\documentclass[conference,10pt]{IEEEtran}
\usepackage{amsmath}
\usepackage{amssymb}
\usepackage{amsfonts}
\usepackage{graphicx}
\usepackage{enumerate}
\usepackage{bbm}
\usepackage{subfig}
\usepackage{mathtools} 
\usepackage{cite}
\usepackage{url}
\usepackage{array}
\usepackage{verbatim}
\usepackage{comment}
\usepackage{stfloats}
\usepackage[draft]{hyperref}

\usepackage{hhline}

\usepackage[font=small,skip=0pt]{caption}
\usepackage{color}
\definecolor{red}{rgb}{1,0,0}
\definecolor{blue}{rgb}{0,0,1}

\IEEEoverridecommandlockouts
\begin{document}
\title{Comparison of Limited Feedback Schemes for NOMA Transmission in mmWave Drone Networks}
\author{
\IEEEauthorblockA{Nadisanka~Rupasinghe\IEEEauthorrefmark{1}, Yavuz~Yap{\i}c{\i}\IEEEauthorrefmark{1},
\. {I}smail~G\"uven\c{c}\IEEEauthorrefmark{1} and Yuichi Kakishima\IEEEauthorrefmark{2}
\IEEEauthorblockA{\IEEEauthorrefmark{1}Department of Electrical and Computer Engineering, North Carolina State University, Raleigh, NC}
\IEEEauthorblockA{\IEEEauthorrefmark{2}DOCOMO Innovations Inc., Palo Alto, CA}
{\tt \{rprupasi, yyapici, iguvenc\}@ncsu.edu,  kakishima@docomoinnovations.com}}%
\thanks{This research was supported in part by the U.S. National Science Foundation under the grant CNS-1618692.}
}
\maketitle

\begin{abstract}

Introducing non-orthogonal multiple access (NOMA) transmission to an unmanned aerial vehicle (UAV) based communication network is a promising solution to enhance its spectral efficiency. However, for realistic deployment of such a network, identifying a practical user feedback scheme for NOMA is essential. In this paper, considering two practical feedback schemes we introduce NOMA transmission to UAVs acting as aerial base stations (BS) to provide coverage at a large stadium. In particular, a UAV-BS generates directional beams, and multiple users are served simultaneously within the same beam employing NOMA transmission. Each user is considered to have a target rate based on its quality of service (QoS) requirements. In order to relieve the burden of tracking and feeding back full channel state information, we consider two limited feedback schemes as practical alternatives: 1) user distance, and 2) user angle with respect to beamforming direction under different user region geometries. Our evaluation results show that NOMA with limited feedback can provide better sum rates compared to its orthogonal counterpart. Further, based on the geometry of user region we identify that there is an optimal feedback scheme and user ordering criteria for NOMA transmission which can maximize sum rates.



\end{abstract}

\begin{IEEEkeywords}
5G, drone, HPPP, mmWave, non-orthogonal multiple access (NOMA), stadium, UAV.
\end{IEEEkeywords}

\section{Introduction}

Unmanned aerial vehicles (UAVs) serving as aerial base stations (BSs) is emerging as a cost-effective and efficient solution for providing rapid on-demand connectivity during temporary events and after disasters~\cite{Zeng2016UAVOpportunities,merwaday2016improved,
UAV_NOMA_Asilomar, NadisankaTCoM_arXiv}. However, due to the limited energy resources on board of a UAV, achieving higher spectral efficiency (SE) is of paramount importance to reap maximum benefits from UAV based communication networks. In this regard, integrating non-orthogonal multiple access (NOMA) to UAV-BSs can be an effective solution to improve SE  \cite{Docomo_NOMA}. In contrast to the conventional orthogonal multiple access (OMA) schemes (e.g., time-division multiple access (TDMA)), NOMA simultaneously serves multiple users in the same time, frequency, code or space resources in a non-orthogonal fashion by considering power domain for multiple access. Hence, low-power UAV-BSs can serve multiple users simultaneously with NOMA using the same resources while enhancing the achievable SE. 

Use of NOMA techniques to improve SE has been studied extensively in the literature in a broader context. In particular, NOMA with multi-antenna transmission techniques is recently receiving higher attention \cite{Ding16MIMO_NOMA,Ding16Schober_MIMO_NOMA,Ding17PoorRandBeamforming, Ding16MIMO_IoT,UAV_NOMA_Asilomar,NadisankaTCoM_arXiv, Zeng17MIMO_NOMA_Capacity_Comp}. In \cite{Ding16MIMO_NOMA} multiple-input-multiple-output (MIMO) techniques are introduced to NOMA transmission along with user pairing and power allocation strategies to enhance MIMO-NOMA performance over MIMO-OMA. A general MIMO-NOMA framework applicable to both downlink (DL) and uplink (UL) transmission is proposed in \cite{Ding16Schober_MIMO_NOMA} by considering signal alignment concepts. A random beamforming approach for millimeter (mmWave) NOMA networks is proposed in \cite{Ding17PoorRandBeamforming}. In that, for user ordering, full channel state information (CSI) of users which depend on the angle offset between the randomly generated base station (BS) beam, user distances and small scale fading are considered. Two users are then served simultaneously within a single BS beam by employing NOMA techniques.


A UAV based mobile cloud computing system is proposed in \cite{7932157} where UAVs offer computation offloading opportunities to mobile stations (MS) with limited local processing capabilities. In that, just for offloading purposes between a UAV and the MSs, NOMA is proposed as one viable solution. In our earlier work \cite{UAV_NOMA_Asilomar, NadisankaTCoM_arXiv}, NOMA transmission is introduced to UAVs acting as aerial BSs to provide coverage over a stadium or a concert scenario. In particular, leveraging multi-antenna techniques a UAV-BS generates directional beams, and multiple users are served within the same beam employing NOMA transmission. In \cite{UAV_NOMA_Asilomar} we assume the availability of full CSI feedback whereas in \cite{NadisankaTCoM_arXiv} the availability of only user distance information is assumed as a practical feedback scheme for NOMA formulation.

In this paper, we consider a similar scenario as in \cite{UAV_NOMA_Asilomar,NadisankaTCoM_arXiv}, where a UAV-BS is employed to provide broadband connectivity over a densely packed user area in a stadium. Then,  NOMA transmission along with multi-antenna transmission is introduced to improve the SE. In order to relieve the burden owing to tracking and feeding back full CSI used by the NOMA transmitter for user scheduling and power allocation, we consider two limited feedback schemes as practical alternatives. In particular, 1) user distance (as in~\cite{NadisankaTCoM_arXiv}), and 2) user angle information, are considered as available user feedback which provide a measure of user channel quality. We show that, depending on the geometry of the area covered by UAV-BS, these limited feedback schemes can provide comparatively different sum rate performances. In particular, numerical results verify that angle as the limited feedback is significantly superior to distance feedback whenever users are more distinguishable by their respective angles (i.e., when the horizontal footprint of UAV-BS beam is sufficiently wider). Further, considering two user ordering criteria with angle feedback (Fej\'er kernel, and absolute angle) achievable sum rate performance under different geometries and NOMA user pairs are evaluated.

\section{System Model} \label{sec:Sys_Model}

\begin{figure}[!t]
\begin{center}
\includegraphics[width=0.4\textwidth]{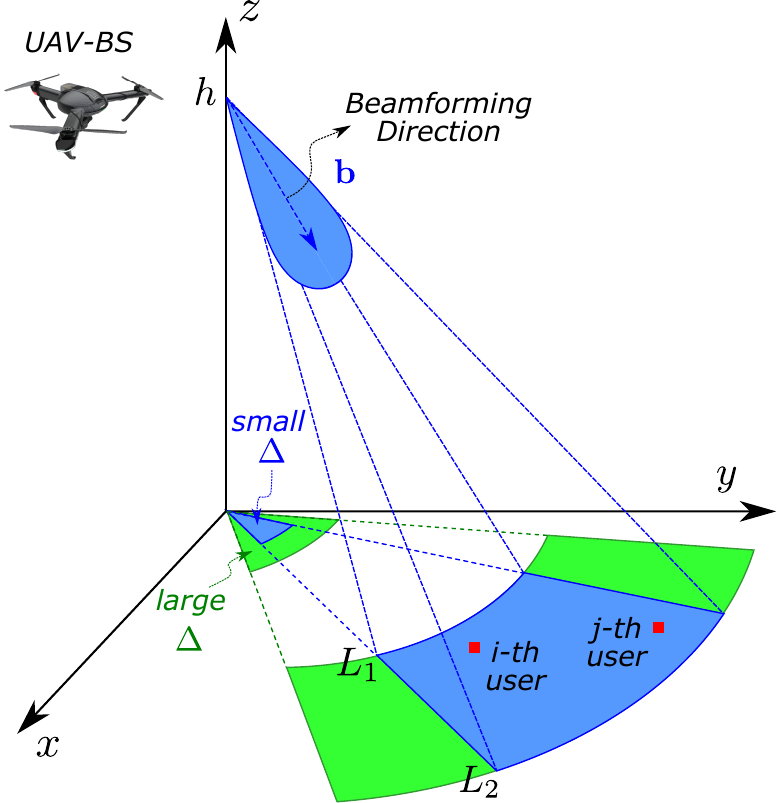}
\end{center}
\caption{System scenario where NOMA transmission serves multiple users simultaneously in a single DL beam (assuming various values for the horizontal angle $\Delta$).}
\label{fig:footprint}
\vspace{0.25 em}
\end{figure} We consider a mmWave-NOMA transmission scenario where a single UAV-BS equipped with an $M$~element uniform linear array (ULA) is serving single-antenna users in the DL. We assume that all these users lie inside a specific \emph{user region} as shown in Fig.~\ref{fig:footprint}. A 3-dimensional (3D) beam is generated by the UAV-BS entirely covering this user region. The set of users within the user region can be represented by the set $\mathcal{N}_{\rm U} = \{1,2,\ldots K\}$ with number of users within the user region being equal to $|\mathcal{N}_{\rm U}|$. The user region is identified by inner-radius $L_1$, outer-radius $L_2$, and $\Delta$, which is the fixed angle within the projection of horizontal beamwidth of the antenna pattern on the $xy$-plane, as shown in Fig.~\ref{fig:footprint}. Note that it is possible to reasonably model various different hot spot scenarios such as a stadium, concert hall, traffic jam, and urban canyon by modifying these control parameters.

\subsection{User Distribution and mmWave Channel Model}
We assume that mobile users are uniformly randomly distributed within the user region following a homogeneous Poisson point process (HPPP) with density $\lambda$. Hence, the number of users in the specified user region is Poisson distributed such that $\textrm{P}(K \textrm{ users in the user region})\,{=}\, \frac{\mu^K e^{{-}\mu}}{K!} $ with $\mu\,{=}\,(L_2^2\,{-}\,L_1^2)\frac{\Delta}{2} \lambda$.

 The channel $\textbf{h}_k$ between the $k$-th user and the UAV-BS is \begin{align} \label{k_UE_original_channel}
\textbf{h}_k = \sqrt{M} \sum \limits _{p=1}^{N_{\rm P}} \frac{\alpha_{k,p} \textbf{a}(\theta_{k,p})}{\sqrt{\textrm{PL}\left(\sqrt{d_k^2 + h^2}\right)}},
\end{align} where $N_{\rm P}$, $h$, $d_k$, $\alpha_{k,p}$ and $\theta_{k,p}$ represent the number of multi-paths, UAV-BS hovering altitude, horizontal distance between $k$-th user and UAV-BS, gain of the $p$-th path which is complex Gaussian distributed with $\mathcal{CN}(0,1)$, and angle-of-departure (AoD) of the $p$-th path, respectively. $\textbf{a}(\theta_{k,p})$ is the steering vector corresponding to AoD $\theta_{k,p}$ given as,
\begin{align*}
\textbf{a}\left( \theta_{k,p} \right) = \left[ 1 \;\; e^{-j2\pi \frac{D}{\lambda}\sin\left( \theta_{k,p}\right) } \; \dots \; e^{-j2\pi \frac{D}{\lambda}\sin\left( \theta_{k,p} \right)\left( M-1\right) } \right]^{\rm T} ,
\end{align*} where $D$ is the antenna spacing of ULA, and $\lambda$ is the wavelength. The path loss (PL) between $k$-th mobile user and the UAV-BS is captured by $\textrm{PL}\left(\sqrt{d_k^2 + h^2}\right)$. Without any loss of generality, we assume that all the users have line-of-sight (LoS) paths since UAV-BS is hovering at relatively high altitudes and the probability of having scatterers around UAV-BS is very small. Furthermore, as discussed in \cite{Ding17PoorRandBeamforming, Lee16mmWaveChannel}, the effect of the LoS link is dominant compared to the Non-LoS (NLoS) links for mmWave frequency bands. Hence, it is reasonable to assume a single LoS path for the channel under consideration, and \eqref{k_UE_original_channel} accordingly becomes
\begin{align} \label{k_UE_modified_LoS_channel}
\textbf{h}_k = \sqrt{M} \frac{\alpha_k \textbf{a}(\theta_k)}{\sqrt{\textrm{PL}\left(\sqrt{d_k^2 + h^2}\right)}},
\end{align}
where $\theta_{k}$ is AoD of the LoS path.

\section{NOMA for UAV-BS Downlink} \label{Sec:NOMA_Transmission}
In this section, we consider NOMA transmission to serve multiple users simultaneously in the DL considering single UAV beam as 
in Fig.~\ref{fig:footprint}. Assuming that each user has its own QoS based target rate, we evaluate respective sum rates to investigate conditions to serve each user at least at its target rate. As considered in \cite{UAV_NOMA_Asilomar,Ding17PoorRandBeamforming}, we assume UAV-BS generates a beam $\textbf{b}$ in azimuth AoD $\overline{\theta}$.  Here, $\overline{\theta}$  can take values either from $[0{,}\,2\pi]$, or a subset of it. 
Note that the full coverage of the entire environment can be achieved by choosing values for $\overline{\theta}$ from its support either sequentially or randomly over time.

Following the convention of~\cite{Ding17PoorRandBeamforming} and as shown in \cite{NadisankaTCoM_arXiv}, we assume that $\Delta$ is small, i.e., $\Delta\,{\rightarrow}\,0$, which results in a small angular offset such that $|\overline{\theta}\,{-}\,\theta_k|\,{\rightarrow}\,0$. Assuming an adequate coordinate system, small angular offset always implies small individual angles such that $\sin\overline{\theta}\,{\rightarrow}\,\overline{\theta}$ and $\sin\theta_k\,{\rightarrow}\,\theta_k$. Using \eqref{k_UE_original_channel}, the effective channel gain of user $k\,{\in}\,\mathcal{N}_\textrm{U}$ for a particular beamforming direction $\overline{\theta}$ is therefore given by
\begin{align} \label{eq:Eff_channel_gain}
|\textbf{h}_k^{\rm H}\textbf{b}|^2 &\approx \frac{|\alpha_k|^2}{M \times\textrm{PL}\left(\sqrt{d_k^2 + h^2}\right)}
 \left| \frac{ \sin \left( \frac{\pi M(\overline{\theta} - \theta_k)}{2} \right)}{  \sin \left( \frac{\pi (\overline{\theta} - \theta_k)}{2} \right)}\right|^2 ,
\nonumber \\
 &= \frac{|\alpha_k|^2}{\textrm{PL}\left(\sqrt{d_k^2 + h^2}\right)} {\rm F}_M(\overline{\theta} - \theta_k),
\end{align} where ${\rm F}_M(\cdot)$ is called the Fej\'er kernel \cite{Ding17PoorRandBeamforming}.
\subsection{Outage Probabilities and Sum Rates} \label{sec:Outage_Prob_Sum_Rates}

Deferring the discussion of user ordering strategies to the next section, we assume without any loss of generality that the users in set $\mathcal{N}_{\rm U}$ are already indexed from the best to the worst channel quality under a given criterion. Defining $\beta_k$ to be the power allocation coefficient of $k$-th user, we therefore have $\beta_1\,{\leq}\,\dots\,{\leq}\,\beta_K$ such that $\sum \limits_{k{=}1}^{K} \beta_k^2\,{=}\,1$. The transmitted signal $\textbf{x}$ is then generated by superposition coding as,
\begin{align}
\textbf{x} = \sqrt{P_{\rm Tx}}\textbf{b}\sum \limits_{k = 1}^{K} \beta_k s_{k}, \end{align} where $P_{\rm Tx}$ and $s_{k}$ are the total DL transmit power and $k$-th user's message, respectively. The received signal at the $k$-th user is then given as
\begin{align} \label{eq:k-th_user_Rx_signal}
y_{k}= \textbf{h}_{k}^{\rm H} \textbf{x} +  v_k = \sqrt{P_{\rm Tx}}\textbf{h}_{k}^{\rm H} \textbf{b}\sum \limits_{k = 1}^{K} \beta_k s_{k} + v_k,
\end{align}
where $v_k$ is complex Gaussian noise with $\mathcal{CN}(0,N_0)$.

At the receiver, each user first decodes messages of all weaker users (allocated with larger power) sequentially in the presence of stronger users' messages (allocated with smaller power). Those decoded messages are then subtracted from the received signal in \eqref{eq:k-th_user_Rx_signal}, and each user decodes its own message treating the stronger users' messages as noise. This overall decoding process is known as successive interference cancellation (SIC), and the respective SINR at $k$-th user while decoding $m$-th user message is
\begin{align} \label{eq:SINR_mk_th_user}
\textrm{SINR}_{m{\rightarrow}k} = \frac{P_{\rm Tx}|\textbf{h}_{k}^{\rm H}\textbf{b}|^2 \beta_{m}^2}{P_{\rm Tx} \sum \limits_{l = 1}^{m-1}|\textbf{h}_{k}^{\rm H}\textbf{b}|^2 \beta_{l}^2 + N_0},
\end{align}
where $k+1 \,{\leq}\, m \,{\leq}\, K$. Assuming that all interfering messages of weaker users are decoded accurately (requires the instantaneous rate associated with decoding any of these weaker users' messages to be larger than the respective target rate of that user), $k$-th user has the following SINR while decoding its own message:
\begin{align} \label{eq:SINR_k_th_user}
{\rm SINR}_{k} =  \frac{P_{\rm Tx}|\textbf{h}_{k}^{\rm H}\textbf{b}|^2 \beta_{k}^2}{ \left( 1- \delta_{k1} \right) P_{\rm Tx}  \sum \limits_{l = 1}^{k-1}|\textbf{h}_{k}^{\rm H}\textbf{b}|^2 \beta_{l}^2 + N_0}.
\end{align}
Here, $\delta_{k1}$ is the Kronecker delta function taking $1$ if $k\,{=}\,1$, and $0$ otherwise. Defining the instantaneous rates associated with \eqref{eq:SINR_mk_th_user} and \eqref{eq:SINR_k_th_user} to be $R_{m{\rightarrow}k}\,{=}\,\log_2 \left( 1\,{+}\,\textrm{SINR}_{m{\rightarrow}k}\right)$ and $R_{k}\,{=}\,\log_2 \left( 1\,{+}\,\textrm{SINR}_{k}\right)$, respectively, the outage probability of $k$-th NOMA user is given as \begin{align}
\textrm{P}_{k}^{o} & = 1 - \textrm{P} \left( \textrm{SINR}_{k+1{\rightarrow}k} > \epsilon_{k+1}, \dots , \textrm{SINR}_{K{\rightarrow}k} > \epsilon_{K}, \right.
\nonumber \\
& \hspace{12em} \left. \textrm{SINR}_{k} > \epsilon_k | \, \mathcal{S}_K \right) \label{eq:Outage_k_th_user_2},
\end{align}
where $\overline{R}_k$, $k\, \in \mathcal{N}_{\rm U}$ is the QoS based target rate for $k$-th user and $\epsilon_k\,{=}\,2^{\overline{R}_k}\,{-}\,1$. Note that \eqref{eq:Outage_k_th_user_2} is defined for the set $\mathcal{S}_K$ describing the given condition on $K$ which involves a range of integers such that $\mathcal{S}_K\,{:}\,\left\lbrace K \,|\, j\,{\leq}\, K \,{<}\, i \right\rbrace$, where $i,\, j \in \mathbb{Z}^+$.


The respective \textit{outage} sum rates are then computed as the weighted sum of target rates, where each target rate is weighted by its non-outage probability, and are given as
\begin{align} \label{eq:sum_rate_NOMA_setK}
R^{\textrm{NOMA}} = \sum \limits_{k = 1}^{K}  (1- \textrm{P}_{k}^{o}) \overline{R}_k,
\end{align} where $K\,{\in}\,\mathcal{S}_K$. Note that whenever we have $K\,{=}\,1$, single user transmission is employed where the full time-frequency resources and transmit power are allocated to the scheduled user. For performance comparison, we also consider OMA outage sum rates as follows
\begin{align}
R^{\textrm{OMA}} = \sum \limits_{k = 1}^{K}  (1- \tilde{\textrm{P}}_{k}^{o}) \overline{R}_k,
\end{align} with $\tilde{\textrm{P}}_{k}^{o}\,{=}\,\textrm{P}\left(\frac1K \log\left(1{+}P_{\rm Tx}|\textbf{h}_{k}^{\rm H}\textbf{b}|^2/N_0 \right)\,{<}\,\overline{R}_k|\mathcal{S}_K \right)$ \cite{UAV_NOMA_Asilomar,NadisankaTCoM_arXiv}.

\subsection{Limited Feedback Schemes and User Ordering} \label{sec:feedback_noma}

Since NOMA transmitter allocates power to its users based on their channel qualities, it needs to order users according to their effective channel gains given in \eqref{eq:Eff_channel_gain}. This strategy therefore requires users to send appropriate information of their respective channel qualities back to the transmitter. When the underlying channel experiences rapid fluctuations over time, tracking of the full CSI at user terminals becomes cumbersome, and frequently sending this information back to the transmitter increases link overhead. Thus, we consider two types of \emph{limited feedback} schemes based on 1) user distance $d_k$, and 2) user angle, ${\theta}_k$. Both distance $d_k$ and angle $\theta_k$ information change much slowly as compared to full CSI, and are, hence, practical alternatives to full CSI feedback.

Note that the user distance $d_k$ and angle $\theta_k$ appear in the effective channel gain expression of \eqref{eq:Eff_channel_gain} within the individual terms ${\textrm{PL}(\sqrt{d_k^2 \,{+}\, h^2})}$ and ${\rm F}_M(\overline{\theta} \,{-}\, \theta_k)$, respectively. While ${\textrm{PL}}(\cdot)$ is a monotonic function of $d_k$, ${\rm F}_M(\cdot)$ is not monotonically varying with $\theta_k$. Hence, the respective optimal ordering strategies based on limited feedback schemes are defined as
\begin{align}
\textrm{Distance:} & \quad d_1 \leq d_2 \leq \dots \leq d_K \, ,\label{eq:distance_ordering} \\
\textrm{Fej\'er Kernel:} & \quad {\rm F}_M(\theta_1) \geq {\rm F}_M(\theta_2) \geq \dots \geq {\rm F}_M(\theta_K) \, ,\label{eq:Fejer_ordering}
\end{align}
where both these schemes guarantee user ordering from best to worst channel quality, and therefore align with the formulations in Section~\ref{sec:Outage_Prob_Sum_Rates}. Although ${\rm F}_M(\cdot)$ is not a monotonic function of $\theta_k$, we will also consider the following suboptimal ordering scheme
\begin{align}
\textrm{Angle: } \quad \tilde{\theta}_1 \leq \tilde{\theta}_2 \leq \dots \leq \tilde{\theta}_K \,, \quad\quad\quad\quad\quad \label{eq:angle_ordering}
\end{align}
where $\tilde{\theta}_k$ is the absolute angle defined as $\tilde{\theta}_k\,{=}\,|\bar{\theta}{-}\theta_k|$. Note that, \eqref{eq:angle_ordering} also satisfies the ordering of channel qualities from best to worst, as in \eqref{eq:distance_ordering}-\eqref{eq:Fejer_ordering}.

In the next section, we evaluate achievable outage sum rates based on these feedback schemes and ordering strategies under different user region geometries. In particular, we consider only $i$-th and $j$-th users in our investigations, though results can be generalized to more than 2 NOMA users as well.


\section{Numerical Results} \label{sec:Numerical_results}
In this section, we present the impact of various limited feedback schemes and user ordering strategies on NOMA and OMA outage sum rates through Monte Carlo simulations. Considering Fig.~\ref{fig:footprint}, we assume that $L_{2}\,{=}\,100$~m, $L_{1}\,{=}\,85$~m, $\Delta\,{\in}\,\left\lbrace 1^{\circ},5^{\circ}\right\rbrace$, $\bar{\theta}\,{=}\,0^{\circ}$, and $M\,{=}\,100$. User distribution is based on HPPP with $\lambda\,{=}\,1$, and user target rates are $\overline{R}_j\,{=}\,6$~bits per channel use (BPCU), $\overline{R}_i\,{=}\,0.5$~BPCU. Power allocation ratios are $\beta_j^2\,{=}\,0.25$ and $\beta_i^2\,{=}\,0.75$, and the noise variance is $N_0\,{=}\,-35$~dBm. The path-loss model is assumed to be $\textrm{PL}(\sqrt{d_k^2 + h^2}) \,{=}\, 1 {+} \left(\sqrt{d_k^2 + h^2}\right)^{\gamma}$ with $\gamma\,{=}\,2$ \cite{Ding17PoorRandBeamforming,UAV_NOMA_Asilomar}, and the UAV-BS altitude is $h\,{\in}\,[10,150]$~m.

\begin{figure}[!t]
\centering
\includegraphics[width=0.5\textwidth]{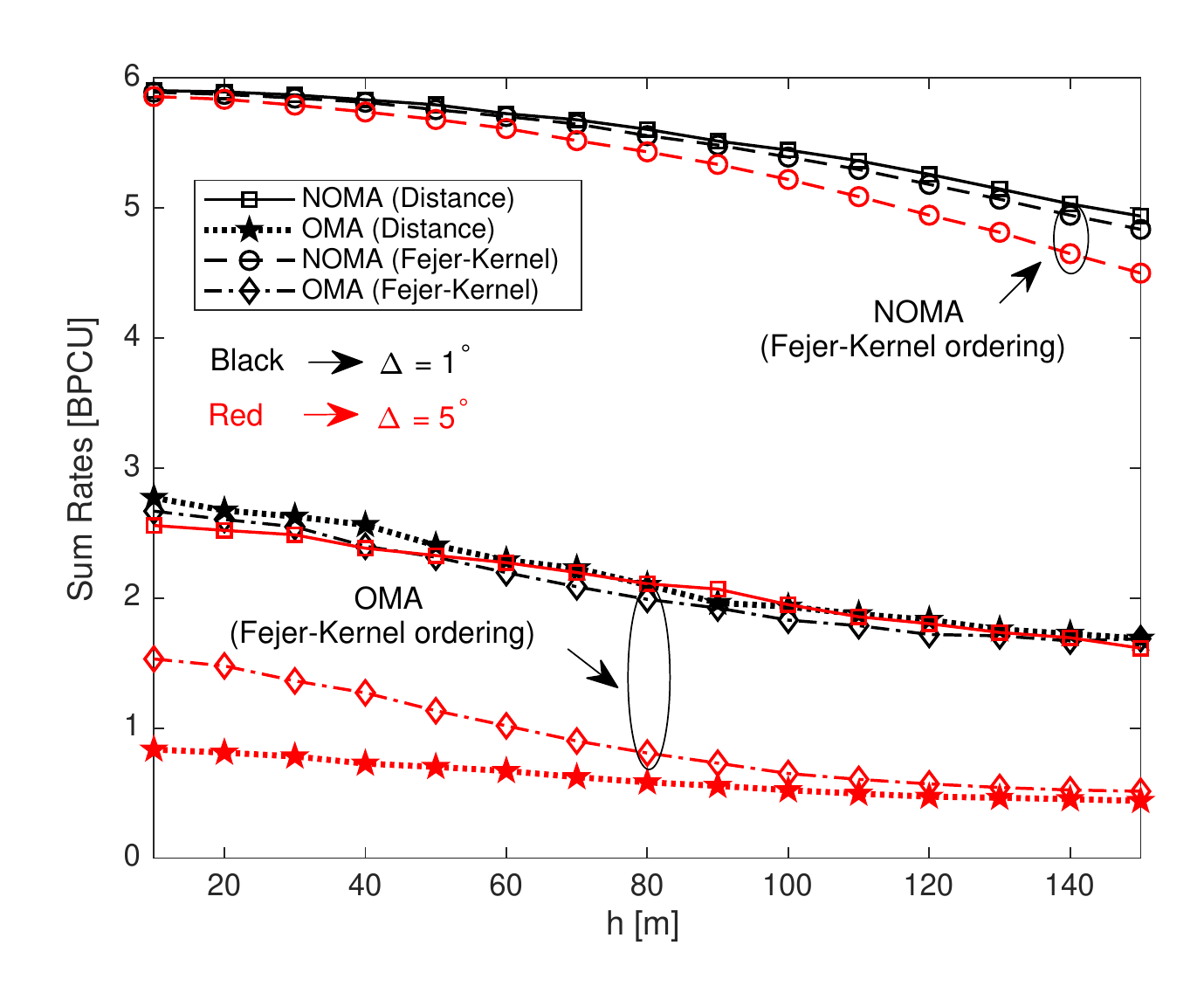}\vspace{0mm}
\caption{Sum rates for NOMA and OMA for distance and Fej\'er kernel based ordering with $i\,{=}\,25$, $j\,{=}\,20$, and $P_{\rm Tx}\,{=}\,20$~dBm.}
\label{fig:sumrate_distance_fejer_NOMA_OMA_j20_i25}
\end{figure}

In Fig.~\ref{fig:sumrate_distance_fejer_NOMA_OMA_j20_i25}, we present outage sum rates of OMA and NOMA along with varying altitudes, where we consider distance and Fej\'er kernel based ordering schemes with $i\,{=}\,25$, $j\,{=}\,20$, $\Delta\,{\in}\,\left\lbrace 1^{\circ},5^{\circ}\right\rbrace$, and $P_{\rm Tx}\,{=}\,20$~dBm. We observe that outage sum rate performance of NOMA outperforms that of OMA for all conditions under consideration. In addition, although both ordering schemes produce very similar performance for a relatively narrow horizontal angle of $\Delta\,{=}\,1^{\circ}$, the Fej\'er kernel based ordering achieves much better rate performance for a wider horizontal angle of $\Delta\,{=}\,5^{\circ}$ while the distance based ordering results in highly degraded sum rate performance.

\begin{figure}[!t]
\centering
\includegraphics[width=0.5\textwidth]{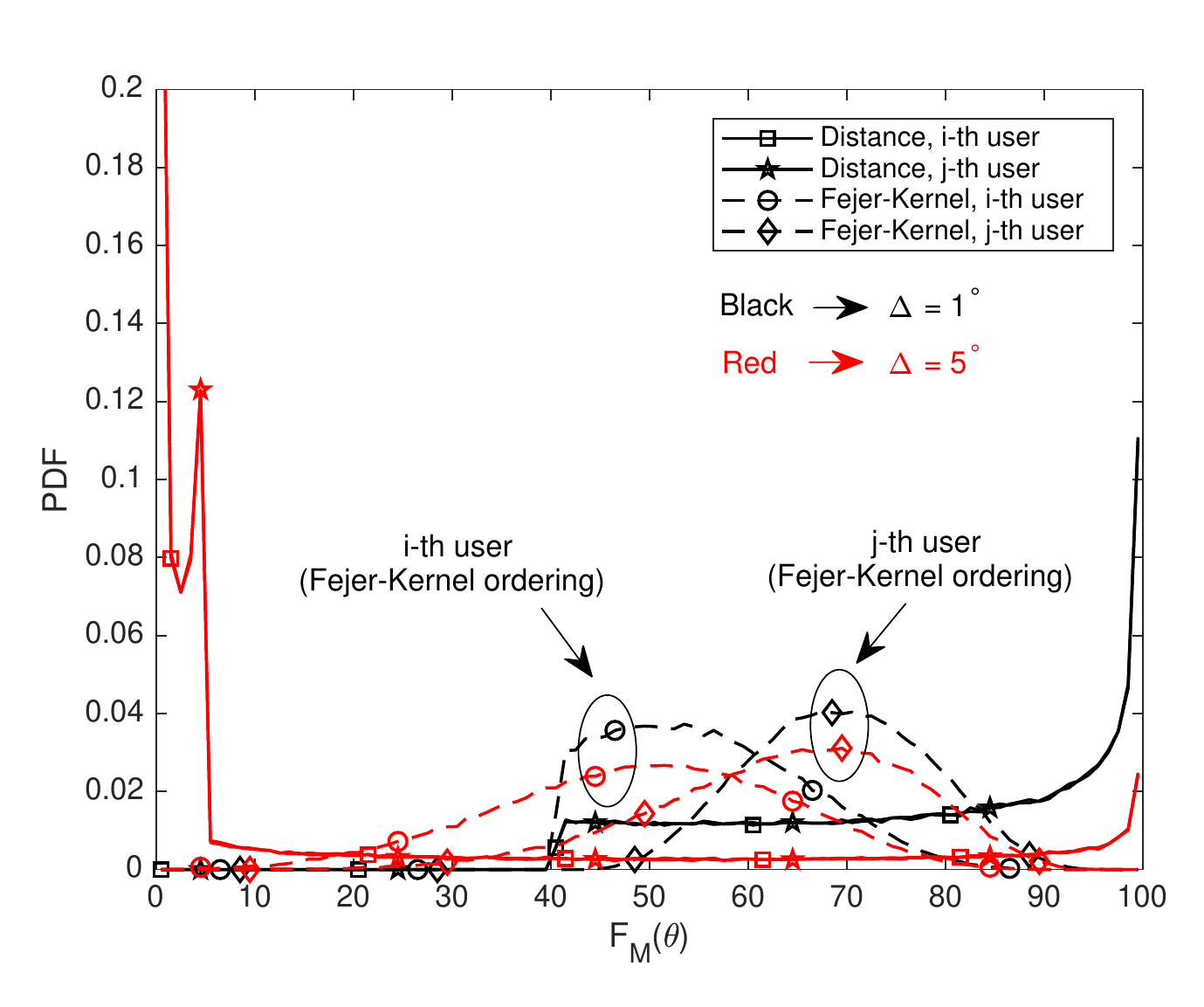}\vspace{-1mm}
\caption{PDF of Fej\'er kernel under distance and Fej\'er kernel based user ordering schemes with $i\,{=}\,25$, $j\,{=}\,20$.}
\label{fig:PDF_Fejer_Kernel}
\end{figure}

In order to further elaborate the impact of distance and Fej\'er kernel based user ordering criteria on rate performance, we plot the PDF of Fej\'er kernel in Fig.~\ref{fig:PDF_Fejer_Kernel} under both ordering schemes. We observe that Fej\'er kernel takes relatively smaller values for both ordering schemes when the horizontal angle becomes wider, i.e., ${\rm F}_M(\theta) \,{\in}\, [40,100]$ when $\Delta\,{=}\,1^{\circ}$ whereas ${\rm F}_M(\theta) \,{\in}\, [0,100]$ when $\Delta\,{=}\,5^{\circ}$, and that the possibility of ${\rm F}_M(\theta)$ taking very small values, i.e., ${\rm F}_M(\theta) \,{\leq}\, 7$, is very likely when the distance based ordering is adopted. As a result, the effective channel gain in \eqref{eq:Eff_channel_gain} is severely affected by small ${\rm F}_M(\theta)$ values for wider horizontal angles and especially with distance based ordering, degrading sum rate performance.

\begin{figure}[!t]
\centering
\includegraphics[width=0.5\textwidth]{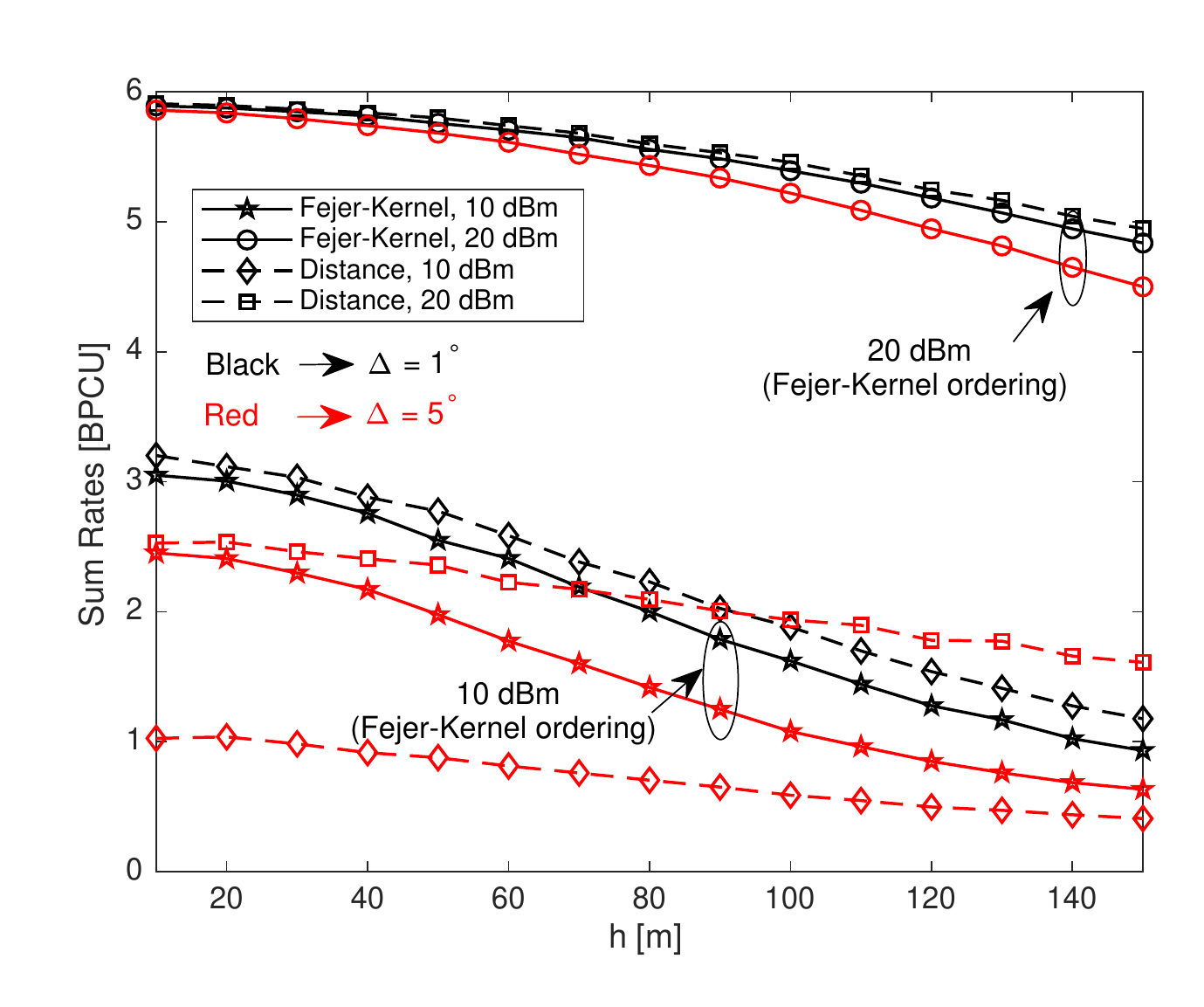}\vspace{-1mm}
\caption{Sum rates for NOMA under distance and Fej\'er kernel based user ordering with $i\,{=}\,25$, $j\,{=}\,20$, and $P_{\rm Tx}\,{=}\,\{10,20\}$~dBm.}
\label{fig:sumrate_distance_fejer_comparison_j20_i25}
\end{figure}

In Fig.~\ref{fig:sumrate_distance_fejer_comparison_j20_i25}, the effect of DL power budget on NOMA sum rates is investigated under distance and Fej\'er kernel based user ordering schemes with $i\,{=}\,25$, $j\,{=}\,20$, $\Delta\,{\in}\,\left\lbrace 1^{\circ},5^{\circ}\right\rbrace$, and $P_{\rm Tx}\,{=}\,\{10,20\}$~dBm. We observe the performance gap between distance and Fej\'er kernel based ordering schemes become larger for relatively small transmit power of $P_{\rm Tx}\,{=}\,10$~dBm, and that this gap is even more significant for the wide horizontal angle of $\Delta\,{=}\,5^{\circ}$.

\begin{figure}[!t]
\centering
\includegraphics[width=0.5\textwidth]{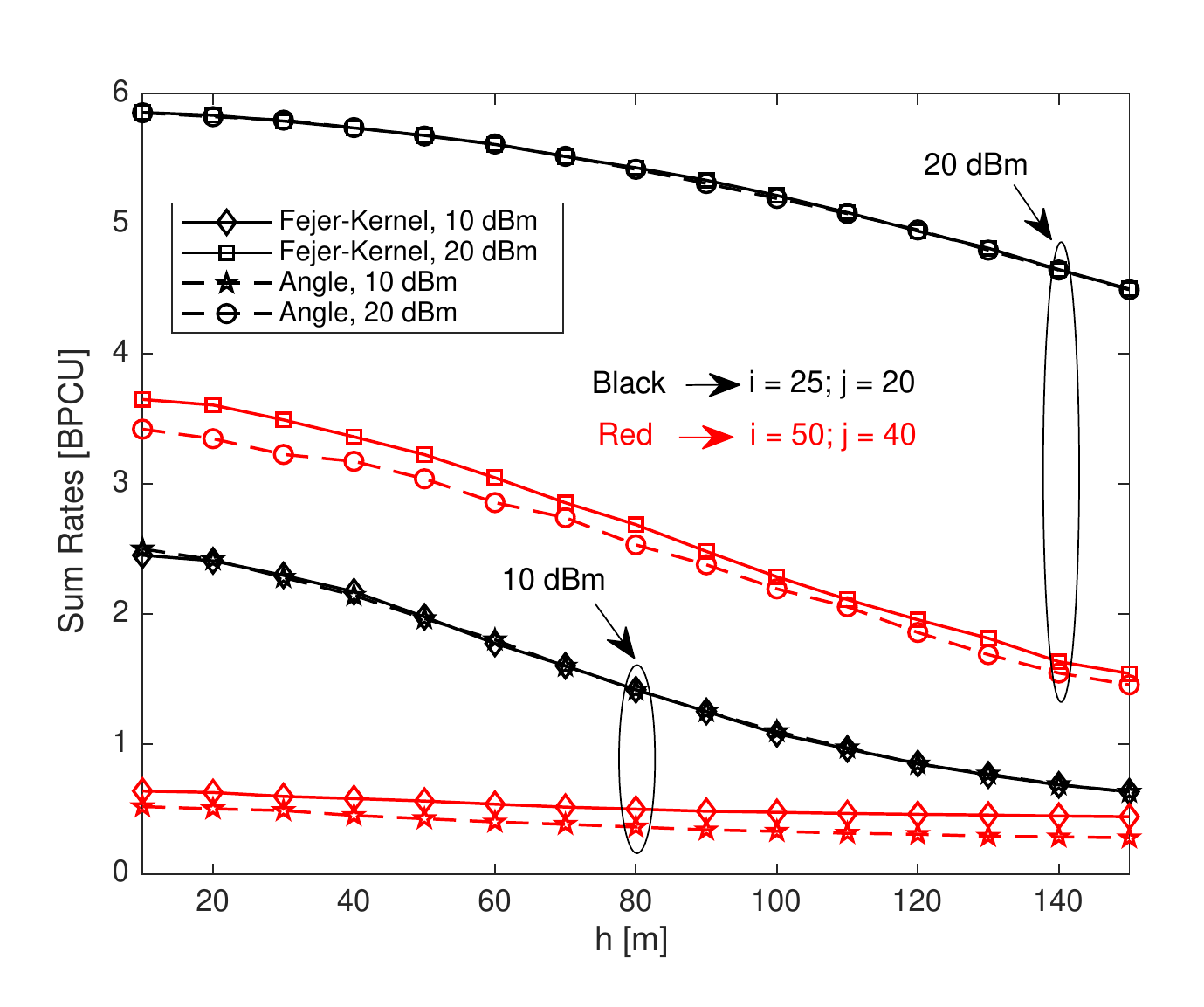}\vspace{-1mm}
\caption{NOMA sum rates comparison for Fej\'er kernel and absolute angle based user ordering, $\Delta\,{=}\,5^{\circ}$.}
\label{fig:sumrate_angle_fejer_comparison_j20_i25}
\end{figure}

\begin{figure}[!t]
\centering
\includegraphics[width=0.5\textwidth]{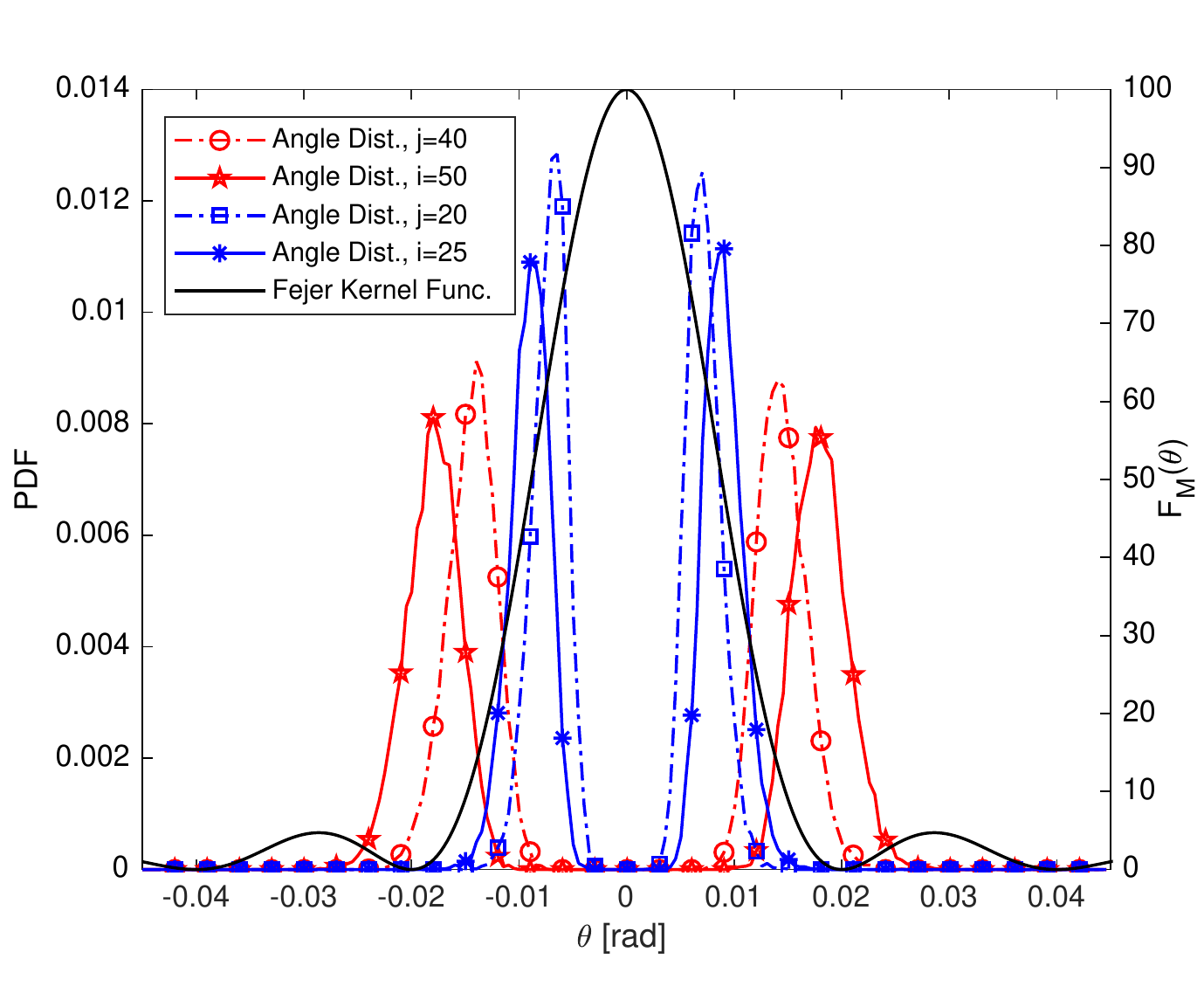}\vspace{-1mm}
\caption{PDFs of angle distribution for absolute angle based user ordering with $\Delta\,{=}\,5^{\circ}$, and $\bar{\theta}\,{=}\,0^{\circ}$ considering user pairs $(i,j)\,{=}\,(50,40)$ and $(i,j)\,{=}\,(25,20)$.}
\label{fig:PDF_Angle_Distribution}
\end{figure}

In Fig.~\ref{fig:sumrate_angle_fejer_comparison_j20_i25}, outage sum rates of NOMA are presented along with varying altitudes, where we consider Fej\'er kernel and absolute angle based ordering criteria with $\Delta\,{=}\,5^{\circ}$ and $P_{\rm Tx}\,{=}\,\left\lbrace 10,20 \right\rbrace$~dBm. In this case, we consider two different users pairs, 1) $i\,{=}\,25$, $j\,{=}\,20$, and 2) $i\,{=}\,50$, $j\,{=}\,40$. We observe that although both ordering criteria perform the same for $i\,{=}\,25$ and $j\,{=}\,20$, sum rate performance of the Fej\'er kernel based ordering is better when $i\,{=}\,50$, $j\,{=}\,40$. To investigate the reason behind this behavior, we plot the Fej\'er kernel function and the PDF of user angle $\theta_k$ in Fig.~\ref{fig:PDF_Angle_Distribution} considering both Fej\'er kernel and absolute angle based user ordering criteria. We observe that the Fej\'er kernel function is decreasing monotonically (for increasing positive angles) within the support of the PDF of ordered $\theta_i$ and $\theta_j$ when $i\,{=}\,25$ and $j\,{=}\,20$. This means that the set of inequalities in the Fej\'er kernel based ordering of \eqref{eq:Fejer_ordering} can be equally represented by those in absolute angle based ordering of \eqref{eq:angle_ordering}, i.e., ${\rm F}_M(\theta_j) \,{\geq}\, {\rm F}_M(\theta_i)$ always corresponds to $\theta_j \,{<}\, \theta_i$, and, hence, both these ordering schemes become equivalent. On the other hand, when we assume the user pair with $i\,{=}\,50$ and $j\,{=}\,40$, the Fej\'er kernel function is non-monotonic within the support of the respective angle PDFs, i.e., it increases for $\theta_k \,{\leq}\, 0.02$~radian and decreases for $\theta_k \,{>}\, 0.02$~radian (along with increasing angle). Therefore, the inequalities in Fej\'er kernel based ordering do not necessarily match those in absolute angle based ordering, i.e., ${\rm F}_M(\theta_j) \,{\geq}\, {\rm F}_M(\theta_i)$ corresponds to either $\theta_i \,{\geq}\, \theta_j$ or $\theta_i \,{<}\, \theta_j$ depending on the particular values of $\theta_i$ and $\theta_j$, which results in observing different sum rate performance for these two ordering criteria with $i\,{=}\,50$ and $j\,{=}\,40$.

\section{Concluding Remarks}\label{sec:conclusion}
In this paper, we introduce NOMA transmission to a UAV-BS flying over a densely packed stadium providing broadband coverage. Two limited feedback schemes are assumed for NOMA which consider user distance and user angle information as practical alternatives to full CSI feedback. Based on these feedback schemes users are then ordered with respect to their distance and angle (considering both Fej\'er kernel and absolute angle) during NOMA formulation.

The numerical results imply that users should be ordered based on a channel quality measure (either distance or angle), on which users become more distinguishable. For instance, whenever the footprint of the UAV beam on the ground is wide enough in horizontal angle, the outage sum rate performance of angle based ordering schemes outperform that of the distance based user ordering. Even though, the Fej\'er kernel based user ordering is the optimal strategy for angle feedback scheme, the performance of user ordering based on absolute angle is also investigated rigorously. In particular, our investigation shows that whenever NOMA user pair has angle support over which the Fej\'er kernel function is monotonically varying, both Fej\'er kernel and absolute angle based ordering strategies provide similar sum rate performance.

\bibliographystyle{IEEEtran}
\bibliography{Doc_Ref}
\end{document}